\documentclass[apj,appendixfloats,numberedappendix]{emulateapj}

\usepackage{amsmath}
\usepackage{wasysym}
\usepackage{rotating}
\usepackage[dvips]{color}
\usepackage{verbatim}
\usepackage{soul}
\usepackage{url}
\usepackage[breaklinks]{hyperref}
\usepackage{breakurl}
\usepackage{lineno}
\usepackage{subfigure}

\slugcomment{Accepted for publication in ApJ.}

\shorttitle{Exploring the Morphology and Origins of the 4C~38.41 Jet}

\shortauthors{Algaba et al.}

\begin{document}

%\linenumbers

\title{Exploring the Morphology and Origins of the 4C~38.41 Jet}

\author{Juan Carlos Algaba$^{1,2,3}$, Bindu Rani$^{4\dagger}$, Sang-Sung Lee$^{3,5}$, Motoki Kino$^{6,7}$, Jongho Park$^{2,8}$, Jae-Young Kim$^9$}
\thanks{$^{\dagger}$NASA Postdoctoral Program (NPP) Fellow}

\affil{
$^1$Department of Physics, Faculty of Science, University of Malaya, 50603 Kuala Lumpur, Malaysia\\
$^2$Department of Physics and Astronomy, Seoul National University, 1 Gwanak-ro, Gwanak-gu, Seoul 08826, Korea\\
$^3$Korea Astronomy \& Space Science Institute, 776, Daedeokdae-ro, Yuseong-gu, Daejeon, Republic of Korea 305-348\\
$^4$NASA Goddard Space Flight Center, Greenbelt, MD 20771, USA\\
$^5$Korea University of Science and Technology, 217 Gajeong-ro, Yuseong-gu, Daejeon 34113, Korea\\
$^6$Kogakuin University of Technology \& Engineering, Academic Support Center, 2665-1 Nakano, Hachioji, Tokyo 192-0015, Japan\\
$^7$National Astronomical Observatory of Japan, 2-21-1 Osawa, Mitaka, Tokyo, 181-8588, Japan\\
$^8$Academia Sinica, Institute of Astronomy and Astrophysics, P.O Box 23141, Taipei 10617, Taiwan\\
$^9$Max-Planck-Institut f\"ur Radioastronomie, Auf dem H\"ugel 69, D-53121 Bonn, Germany\\
}

\begin{abstract}
%The flat spectrum radio quasar 1633+382 (4C~38.41) showed the ejection of VLBI components during the period 2012 March -- 2015 August. Here we study the properties of the innermost jet of this source based on these components and additional VLBI data from the radio monitoring observations of the Boston University VLBI program at 43~GHz. Analysis of the components suggests a semi-parabolic jet geometry with jet radius $R$ following the relation $R\propto r^{0.7}$ with distance $r$ within the Bondi radius, with indications of a jet geometry break towards a conical geometry. This  parabolic to conical transition strongly supports a unification model for the jet geometry for various kinds of AGN.  Brightness temperature falls with distance following $T_B\propto r^{-2.1}$. Combining this information, magnetic field and electron densities are found to fall along the jet as $B\propto r^{-1.5}$ and $n\propto r^{-1.1}$ respectively, suggesting that the magnetic configuration in the jet may be dominated by the poloidal component. Our analysis of the jet structure suggests that the innermost jet regions  do not follow a ballistic trajectory and, instead, match a sinusoidal morphology that may be explained by a binary black hole or Lense-Thirring precession of the accretion disk.
We study the properties of the innermost jet of the flat spectrum radio quasar 1633+382 (4C~38.41) based on VLBI data from the radio monitoring observations of the Boston University VLBI program at 43~GHz. Analysis of the components suggests a semi-parabolic jet geometry with jet radius $R$ following the relation $R\propto r^{0.7}$ with distance $r$, with indications of a jet geometry break towards a conical geometry.  Brightness temperature falls with distance following $T_B\propto r^{-2.1}$. Combining this information, magnetic field and electron densities are found to fall along the jet as $B\propto r^{-1.5}$ and $n\propto r^{-1.1}$ respectively, suggesting that the magnetic configuration in the jet may be dominated by the poloidal component. Our analysis of the jet structure suggests that the innermost jet regions  do not follow a ballistic trajectory and, instead, match a sinusoidal morphology which could be due to jet precession from a helical pattern or Kelvin-Helmholtz instabilities.
\end{abstract}


\section{Introduction}
The source 1633+382 (4C~38.41) is a flat spectrum radio quasar (FSRQ) at a redshift $z=1.813$ \citep{Hewett10}. Strong variability in its radio flux has been observed \citep{SpanglerCotton81,Kuhr81,Seielstad85,Aller92} and superluminal motion with jet velocities up to $394\pm23~\mu$as~yr$^{-1}$ ($30.8\pm1.8~c$) has been detected \citep{Lister19}. 4C~38.41 is also well known to be one of the most powerful $\gamma-$ray extragalactic objects \citep{Abdo09,Acero15}, with prominent $\gamma-$ray outbursts.

The $\gamma-$ray flares have been considered to be connected with interaction with emerging VLBI components \citep{Jorstad11} or with Doppler boosting variations geometrically connected to changes in the viewing angle \citep{Raiteri12}. \cite{PaperI} (hereafter, Paper~I) found that major $\gamma-$ray flares were well matched with similar activity in optical and radio bands, with gamma-ray flares leading radio flares. Considering the shock-in-jet model, it was estimated that the location of the radio emitting regions was of the order of 40~pc from the central engine. \cite{PaperII} (hereafter, Paper~II) showed that several $\gamma-$ray flares coincided with the ejection of respective new very long baseline interferometry (VLBI) components which evolved with predominantly adiabatic losses. The source of activity was considered to be dominated with a particle--dominated region, possibly near but downstream the acceleration and collimation region. 

Although the ejection of components has been extensively analysed, the overall structure of the jet is still not very well studied. Very Large Array (VLA) observations have found that its kiloparsec scale morphology shows a core-dominated triple structure in the north-south direction with an extent of about 12 arcsec \citep{Murphy93}, whereas parsec scales show a misalignment of about $90\degr$ with a single-jet structure detected by the Very Long Baseline Array (VLBA) toward the west.  It has been estimated that the parsec-scale jet is aligned at $\sim1\degr - 3\degr$ to our line of sight \citep{Hovatta09,Liu10}.  

Several authors have recently studied the innermost structure of the jet. In \cite{Pushkarev17} the regions within  the upper 100~mas were fitted with a power law suggesting a conical jet with power index $\epsilon=0.95\pm0.01$, whereas the 5~mas showed an almost parabolic jet with $\epsilon=0.57\pm0.01$, where $\epsilon=1$ for the purely conical and $\epsilon=0.5$ for the purely parabolic cases. Nonetheless, no further discussion was conducted. On the other hand, \cite{Algaba17} found a rather conical geometry with $\epsilon=0.14\pm0.21$ for the innermost jet scales based on a fit of the core sizes considering core shift effect indicated, albeit subjected to sparse and possibly variable data. The evolution along the jet of the physical parameters, such as magnetic field strength or electron density are however not well studied.

In this paper we study the properties of the jet within the first milliarc-seconds from the VLBI core region. In particular, we examine the flux density, brightness temperature and collimation profile. Based on our findings, we additionally investigate the magnetic and electron density profiles, the physical origin of the jet in this source, and possible implications. The contents of the paper is organized as follows: In Section 2,  we summarize the observations and data analysis. In Section 3 we compile our results. In Section 4 we discuss the implications of our results, and in Section 5 we draw our conclusions.

\section{Observations and Data Analysis}

For our analysis, we considered the data from the Boston University (BU) 43~GHz Monitoring program\footnote{\url{https://www.bu.edu/blazars/VLBAproject.html}}. The model-fitted data from \cite{Jorstad17} contains models for epochs spanning from 2007 June 15th to 2013 January 16th. To complement this data, we included the model-fitted data analysed in Paper~II corresponding to observations during the period 2012 March -- 2015 June. In Paper~II we discussed the data processing, including components fitting. Here we briefly sumarize the methodology. We used the Caltech \emph{Difmap} package \citep{Shepherd97} to model--fit the various components of the VLBA data. In addition to the core, we could fit other long-lived components that we identify as C2, C3 and C4. We considered resolution limits as given by \cite{Lobanov05} and filter out non-resolved components. If the component appears to be resolved, we consider their distance relative to the core position, and the uncertainty estimates are given for the size, distance, flux density and position angle as $\sigma_d\sim d/\sqrt{\rm DR}$, $\sigma_r\sim(1/2)\sigma_d$, $\sigma_F\sim\sigma_{rms}\sqrt{\rm DR}$ and $\sigma_{\rm PA}\sim\arctan({\sigma_r/r})$ respectively, for dynamic ranges DR$\gg1$ \citep[see e.g.][]{Fomalont99,Lee08}, where $\sigma_{rms}$ is the map root mean square noise level. When combining the two data sets, we obtained a total of 88 unique epochs from 2007 June 15th to 2015 June 9th.

Due to the core shift effects, the location of the VLBI core does not correspond to that of the central engine. Thus, in order to properly measure the position of the components with respect to the central engine, we need to take this effect into account and consider which is the core shift of this source at 43~GHz. For this, we obtained the core shift measure and core shift at 15.4~GHz from \cite{Pushkarev12} and estimated the core shift at 43~GHz to be $\Delta r=14\pm4$~pc (see also Paper~I). This corresponds to angular scales of $\Delta r=0.06\pm0.02$~mas which, although quite small, are still significant for VLBA observations at 43~GHz, with beam sizes of the order of 0.2~mas, suggesting practical achievable resolutions of the order of 0.04~mas when considering high dynamic ranges \citep[typically 400:1 or higher; see, e.g.,][Paper~II]{Jorstad17} of the VLBA images \citep{Lobanov05}. Thus, even such small core shift is comparable with the scales measured here and will have practical effects in the proper analysis of the jet component positions when studied relative to the central engine. In the following we focus on the results considering the distances with respect to the central engine.%will compare results from the VLBI core and from the central engine perspectives.

\section{Results}

\begin{figure}
\center
\includegraphics[scale=0.45,trim={0cm 0cm 0cm 0cm},clip]{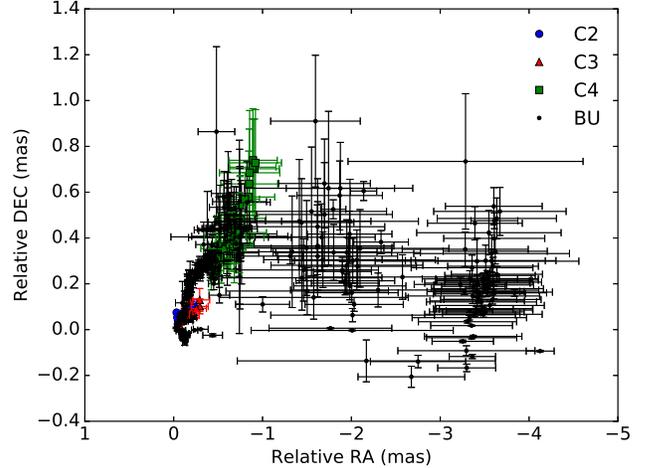}
\caption{Trajectories of the jet components in the RA - DEC plane, where RA$=r \cos(\Theta+ 90)$ and DEC$=r \sin(\Theta + 90)$. Here, $r$ is the radial distance from the core, and $\Theta$ is the position angle of the component with respect to an imaginary north-south line drawn through the map center. Blue circles, red triangles, green squares and black dots indicate components C2, C3, C4 and these from BU Blazar program described in \cite{Jorstad17}, respectively.}
\label{RA-DEC-plane}
\end{figure}

A dedicated study of components distance from the core, including calculations regarding ejection epoch and speed, was already shown in Paper~II. Here we focus on other phenomenological characteristics of the components. In Figure \ref{RA-DEC-plane} we show the trajectories of the jet components in the RA-- DEC plane. The most remarkable feature is an apparent bending of the jet direction. Although the uncertainties are comparatively large at distances $>2$~mas, it is clear that the jet direction does not remain constant and there is a significant bending at a distance of $\sim1$~mas from the core. Indeed, the median value for the jet position angle for distances under $\sim1$~mas is about $\left<PA\right>\sim-59^{\circ}$, whereas for larger distances, $\left<PA\right>\sim-85^{\circ}$.

\begin{figure}
\center
\includegraphics[scale=0.45,trim={0cm 0cm 0cm 0cm},clip]{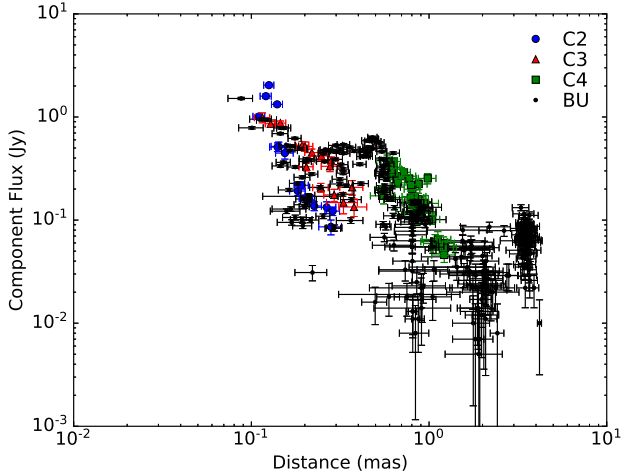}
\caption{Components flux as a function of distance. Top: as observed, considering distances from  the 43~GHz VLBI core. Bottom: considering distances from the central engine, including the core shift at 43~GHz. Color markers and shapes are as Fig. \ref{RA-DEC-plane}.}
\label{Flux}
\end{figure}

Evolution of flux density with distance is shown in Figure \ref{Flux}. Flux density monotonically decreases with distance from the core, with the exception of a slightly larger flux observed in the furthest components (at $\sim3.5$~mas from the core) of the BU data. A plateau in the flux density is also noticeable around $\sim0.4$~mas, although several components appear to have a lower flux densities between 0.2-0.3~mas. It is possible that we are tracing the flux of various different components that intrinsically started with different flux density, leading to a fiducial flux density profile, and in particular, a plateau, along the jet. However, identification of C4 component with that labeled B1 in \cite{Jorstad17} suggests otherwise, as this component seems to have always been below the 1~Jy threshold.

\begin{figure}
\center

\includegraphics[scale=0.45,trim={0cm 0cm 0cm 0cm},clip]{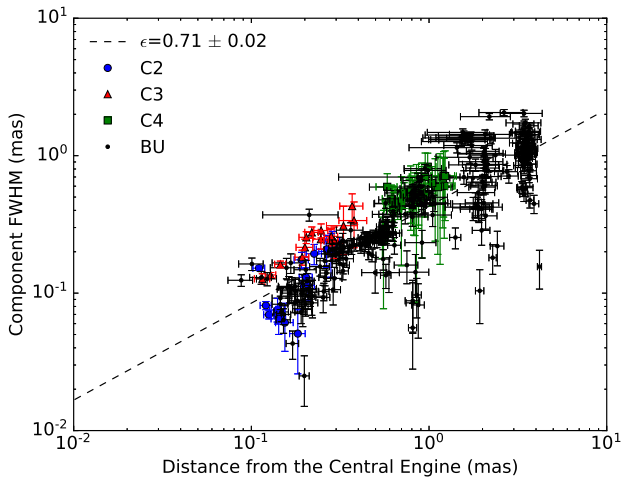}
\caption{Components size as a function of distance. Markers color and shape, and top and bottom panels as in Figure \ref{Flux}.}
\label{Size}
\end{figure}

Figure \ref{Size} shows the deconvolved FWHM of components as a function of distance. The jet component size increases with distance from the core. If we interpret the FWHM of the components with the jet diameter, we can relate the FWHMs as a probe to estimate the jet radius $R$, which we consider to vary with distance $r$ in the form $R\propto r^{\epsilon}$, with $\epsilon=1$ for a purely conical and $\epsilon=0.5$ for a pure parabolic case, respectively \citep{Algaba17}. Due to the large scatter and uncertainties, we considered a Markov-chain Montecarlo (MCMC) method \citep{Foreman-Mackey13} to obtain a more robust fit to the data. A fit including all components indicates $\epsilon=0.62\pm0.01$ considering distances from the VLBI core, or $\epsilon=0.71\pm0.02$ if we consider core shift effects: i.e, from the central engine. This suggests that the jet components in this source follow a quasi--parabolic geometry at the distances probed here.

\begin{figure}
\center
\includegraphics[scale=0.45,trim={0cm 0cm 0cm 0cm},clip]{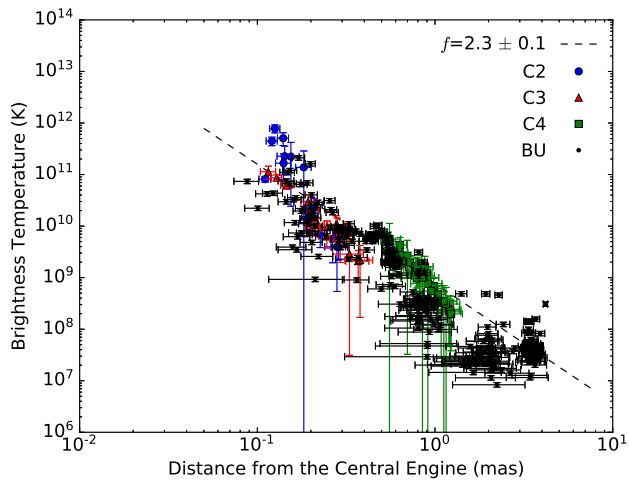}
\caption{Components brightness temperature as a function of distance.  Markers color and shape, and top and bottom panels as in Figure \ref{Flux}.}
\label{Tb}
\end{figure}

The redshift-corrected brightness temperature $T_{B}$ of the bright emission features can be approximated using the following relation \citep{Jorstad05}:
\begin{equation}
T_{B} = 1.22 \times 10^{12} \frac{S_{\rm comp} (1+z)}{{\rm FWHM}^{2} ~\nu^{2}}\, ~ {\rm K},
\end{equation}
where $S_{comp}$ is the component flux density in Jy,  $\nu$ is the observing frequency in GHz, and the FWHM is measured in mas\footnote{Some authors \citep[e.g.][]{Kovalev05} include the factor $(1+z)$ in the numerator, while others \citep[e.g.][]{Hovatta09} include it in the denominator. This will not affect our results since it will appear as a constant factor.}. Table \ref{brightnesstemperature} summarizes the derived components brightness temperature, and Figure \ref{Tb} shows them as a function of distance from the core, together with the brightness temperatures from the BU data \citep{Jorstad17}

Brightness temperatures are of the order of $10^{11}$~K for the innermost components and, similar to the flux density, decrease with distance down to about $10^7$~K in the further components presented here. Evolution of the brightness temperature seems to follow a pattern similar to that of the flux density, with indications of a possible plateau also at 0.3-0.4~mas. A fit of the brightness temperature of the various components as a function of distance in the form $T_B\propto r^{-f}$ indicate $f=2.1\pm0.1$ and $2.3\pm0.1$ when we do not and we do consider the core shift, respectively.

\begin{table}
\begin{center}
\caption{Brightness Temperatures}
\label{brightnesstemperature}
\begin{tabular}{ccccc}
\tableline
\hline
&&\multicolumn{3}{c}{$T_b$ ($\times10^9$ K)}\\
Date &  MJD & C2 & C3 & C4\\
(1)&(2)&(3)&(4)&(5)\\
\tableline
2012 Apr 03	&	56021	&$	 -			$&$	19	\pm	13	$&$	4.4	\pm	1.6	$\\
2012 May 27	&	56074	&$	 -			$&$	 -			$&$	3.2	\pm	2.1	$\\
2012 Jul 05	&	56113	&$	 -			$&$	38	\pm	22	$&$	4.1	\pm	1.9	$\\
2012 Aug 13	&	56153	&$	 -			$&$	478	\pm	130	$&$	3.1	\pm	1.4	$\\
2012 Oct 07	&	56208	&$	 -			$&$	96	\pm	30	$&$	2.5	\pm	1.3	$\\
2012 Oct 19	&	56220	&$	 -			$&$	80	\pm	22	$&$	2.4	\pm	1.5	$\\
2012 Oct 27	&	56228	&$	 -			$&$	58	\pm	18	$&$	2.6	\pm	1.4	$\\
2012 Oct 28	&	56229	&$	 -			$&$	 -			$&$	2.8	\pm	1	$\\
2012 Dec 21	&	56283	&$	 -			$&$	116	\pm	75	$&$	2.2	\pm	2.8	$\\
2013 Jan 15	&	56308	&$	 -			$&$	87	\pm	15	$&$	2.1	\pm	0.7	$\\
2013 Feb 26	&	56350	&$	 -			$&$	62	\pm	13	$&$	2	\pm	0.7	$\\
2013 Apr 17	&	56399	&$	 -			$&$	30	\pm	12	$&$	1.6	\pm	1	$\\
2013 May 31	&	56443	&$	 -			$&$	31	\pm	17	$&$	1.6	\pm	1.3	$\\
2013 Jul 01	&	56474	&$	65	\pm	16	$&$	6	\pm	3	$&$	1.2	\pm	0.5	$\\
2013 Jul 29	&	56502	&$	48	\pm	14	$&$	 -			$&$	1.5	\pm	0.9	$\\
2013 Aug 26	&	56531	&$	81	\pm	24	$&$	5.6	\pm	4	$&$	1.1	\pm	0.8	$\\
2013 Nov 18	&	56615	&$	447	\pm	118	$&$	10.5	\pm	6.1	$&$	1.2	\pm	1	$\\
2013 Dec 16	&	56643	&$	442	\pm	67	$&$	8.1	\pm	3.1	$&$	1	\pm	0.6	$\\
2014 Jan 20	&	56678	&$	780	\pm	123	$&$	9.5	\pm	3.3	$&$	0.7	\pm	0.3	$\\
2014 Feb 25	&	56714	&$	491	\pm	82	$&$	10	\pm	3.1	$&$	0.7	\pm	0.3	$\\
2014 May 04	&	56781	&$	227	\pm	122	$&$	2.1	\pm	1.8	$&$	0.5	\pm	0.6	$\\
2014 Jun 21	&	56829	&$	166	\pm	90	$&$	2.2	\pm	2.3	$&$	0.8	\pm	0.9	$\\
2014 Jul 29	&	56867	&$	222	\pm	145	$&$	2.9	\pm	3.3	$&$	1	\pm	1.3	$\\
2014 Sep 23	&	56924	&$	115	\pm	54	$&$	 -			$&$	1.3	\pm	0.7	$\\
2014 Dec 05	&	56997	&$	22	\pm	8	$&$	 -			$&$	1.2	\pm	0.4	$\\
2015 Apr 12	&	57124	&$	6	\pm	4	$&$	 -			$&$	1.2	\pm	0.7	$\\
2015 May 12	&	57154	&$	4	\pm	4	$&$	 -			$&$	1	\pm	0.7	$\\
2015 Jun 09	&	57182	&$	4	\pm	2	$&$	 -			$&$	1	\pm	0.4	$\\
\tableline
\end{tabular}
\end{center}
\end{table}

\begin{figure}
\center
\includegraphics[scale=0.45,trim={0cm 0cm 0cm 0cm},clip]{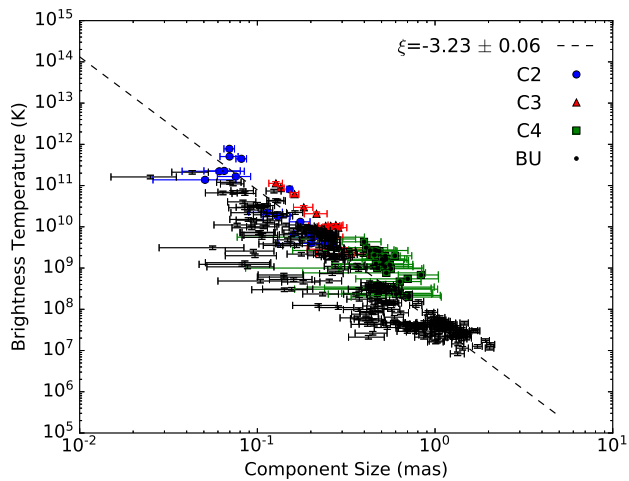}
\caption{Components brightness temperature as a function of component size. Markers color and shape, and top and bottom panels as in Figure \ref{Flux}.}
\label{Tb_vs_size}
\end{figure}

\section{Discussion}

\subsection{Jet Geometry}
In this paper we study the properties of the innermost regions of the 1633+382 jet. As we anticipated in Section 2, the 43~GHz VLBI core distance from the central engine is comparable to the scales of the closest components analyzed here, which implies that its consideration will have a significant impact in the interpretation of our results, as seen in Figures \ref{Flux}, \ref{Size}, and \ref{Tb}. Given that we regard the location of the central engine as the upstream end of the jet derived from the core shift, we will hereafter concentrate our discussion based on analysis taking the core shift into account.

Our results above show that the jet width follows a semi-parabolic geometry, with $\epsilon\sim0.7$, between the pure conical and the parabolic cases. Similar results are obtained using the data from \cite{Jorstad17}, which leads to $\epsilon=0.66\pm0.02$ (Note that their published data spans a different time range than the one discussed here). We find however a discrepancy with the results found in \cite{Algaba17}, where a study based on core sizes related with their respective core--shifted locations suggested a rather conical structure with $\epsilon=1.14\pm0.21$ for this source. This value is clearly larger than the one found here. We note however that the results in \cite{Algaba17} are susceptible to variability effects, as the data used there was not simultaneous. It may also be possible that these two methods are probing different jet regions of 1633+382, since the core size was used in \cite{Algaba17}, whereas we only include optically thin jet structure in this work. Another possibility is that we may be probing even different jet structure, if an unresolved spine--sheath morphology, such as the one seen in M87 \citep{Asada16}, is present in this source; or if, given the small viewing angle, we may be actually looking throughout the jet itself as we approach the regions near the base of the jet, thus affecting the observed geometry.

Semi--parabolic jet geometries have been conclusively  found in other sources such as M87 \citep{Asada12}, NGC6251 \citep{Tseng16}, NGC4261 \citep{Nakahara18} or 1H0323+342 \citep{Hada18}. In these sources, this geometry is found upstream the jet, before a transition towards a conical geometry occurs at around the Bondi Radius, at about few $10^5~r_s$, where $r_s$ is the Schwarzschild radius. For the case of 1633+382, the semi--parabolic geometry is found on scales of a milli-arcsecond, which translate to projected distances of the order of $\lesssim8.5$~pc; or $\lesssim250$~pc deprojected, assuming a viewing angle of $\theta\sim2\degr$ to our line of sight \citep{Hovatta09,Liu10}. If we consider the central engine to host a supermassive black hole with a mass of $M\sim1.32\times10^9M_{\odot}$ \citep{Zamaninasab14}, such distances are of the order of $\lesssim2\times10^6~r_s$. This is slightly larger but of the same order of magnitude than the scales discussed in the literature for the other sources.

A more quantitative estimation of the regions being probed here can be done as follows. The Bondi radius can be obtained from $r_B=2GM/c^2_s=(c/c_s)^2 r_s$, where $G$ is the gravitational constant and the sound speed $c_{\rm s}=\sqrt{\gamma k_{\rm B} T/(\mu m_{\rm p})}$  is a function of gas temperature $T$, $\mu (=0.6)$ is the mean molecular weight, $\gamma (=5/3)$ is the adiabatic index of the accreting gas, and $m_{\rm p}$ is the proton mass. \cite{BednarekKirk95} estimated a maximum temperature of about 0.30~keV, which allows us to estimate $r_B\gtrsim2\times10^6~r_s$. Alternatively, the sphere of gravitational influence (SGI), the radius inside of which the black hole gravitational potential dominates, can be written as $r_{SGI}=GM_{BH}/\sigma^2=(1/2)(c/\sigma)^2 r_s$, where $\sigma$ is the stellar velocity dispersion. Based on the M$-\sigma$ relationship \citep[see e.g.][]{Gebhardt2000,FerrareseMerrit2000,MerritFerrarese2001,McConnell2011}, we can consider $M_{BH}=1.9\times10^8 M_{\odot}(\sigma/{\rm [200 km/s]})^{5.12}$, and evaluate $\sigma\sim290$~km s$^{-1}$. With such velocity dispersion, we estimate $r_{SGI}\sim5.5\times10^5 r_s$.

%It thus seems that both the Bondi radius and the radius of gravitational influence are of the same order of magnitude as the regions downstream the jet measured in our data. This suggests that the jet geometry that our measurements describe here is that of the inner regions, where the gravitational influence of the central engine is still significant compared with that of the environment. If the same phenomenology seen in M87 and NGC6251 also apply to 1633+382, it is then not surprising to expect the observed quasi--parabolic geometry at these regions.

The estimated Bondi radius and the radius of gravitational influence appear to have a similar order of magnitude than that of the regions downstream the jet measured in our data. Although the uncertainties are very large, it is possible that our observations are probing the quasi-parabolic geometry of the regions where the gravitational potential of the black hole is still dominant with respect to the thermal energy of the hot gas in the environment, if the same phenomenology seen in M87 and NGC6251 also applies to 1633+382.

\begin{figure}
\center
\includegraphics[scale=0.52,trim={0.2cm 0cm 0cm 0cm},clip]{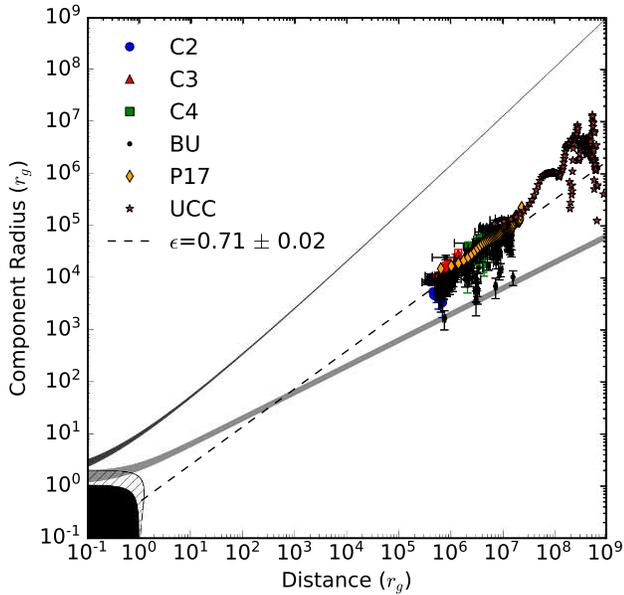}
\caption{Jet radius of 1633+382 including the data from the components discussed here and the 43~GHz BU VLBA Blazar program data, in units of gravitational radii. Markers color and shape are as in Figure \ref{Flux}, with orange diamonds indicating the data from \cite{Pushkarev17} and brown stars the data from the 1.4~GHz VLBA University College Cork Program. The filled black region denotes the black hole (inside the event horizon), while the hatched area represents the ergosphere for the black hole spin parameter a=0.998. Dashed line represents $R\propto r^{0.7}$. %Vertical dotted line represents a very approximate location for the Bondi radius, for reference. 
The light gray area denotes the genuine parabolic streamline ($R\propto r^{1/2}$ at $R\gg r_g$) of the force-free steady jet solution, while the dark gray area denotes the quasi-conical streamline ($R\propto r^{0.97}$ at $R\gg r_g$) of the force-free steady jet solution. In both streamlines, a variation from a=0.5 (upper boundary) to a=0.998 (lower boundary) is considered as a shaded area. Note that all streamlines are anchored at the event horizon in polar coordinates in the Boyer-Lindquist frame, which matches the data coordinates for $R\gg r_g$.}
\label{fig:width_rg}
\end{figure}

In order to investigate this more deeply, we consider the jet geometry in terms of gravitational radii $r_g=GM/c^2$. For 1633+382, 1~mas$\sim4\times10^6~r_g$, considering deprojection effects. In Figure \ref{fig:width_rg} we plot the jet geometry in terms of $r_g$ adding the 43~GHz VLBA-BU Blazar Monitoring Program data as presented in \cite{Jorstad17}, the jet width as in \cite{Pushkarev17}, and the jet width from the 1.4~GHz VLBI observations from the University College Cork program\footnote{\url{http://www.physics.ucc.ie/radiogroup/18-22cm_observations.html}}.  We corrected for the core shift effects at each frequency considered. It seems apparent that the geometry of the jet falls well within the area delimited by the genuine parabolic and the conical streamlines (compare this with Figure 3 of \cite{Algaba17}). 

%While components upstream the jet seem to follow a quasi-parabolic geometry, data from \cite{Pushkarev17} may not follow this extrapolation towards larger radii, and data from University College Cork shows a clear deviation, suggesting that such quasi-parabolic geometry may not the case downstream. Indeed, \cite{Pushkarev17} fitted this region as a geometrically conical jet, with $\epsilon=0.95\pm0.01$. Although more data is required on larger scales both upstream and downstream the jet, it seems that there may be a jet geometry break occurring near the Bondi radius. This phenomenology seems to match the one seen in M87 \citep{Asada12}, NGC6251 \citep{Tseng16} or 3C~273 \citep{Akiyama18}. If this is the case, this would strongly support a unification in the jet geometry aspects for various kinds of AGN.

While components upstream the jet appear to follow a quasi-parabolic geometry, data from \cite{Pushkarev17} may not follow this extrapolation towards larger radii, and data from University College Cork shows a clear deviation, suggesting that such quasi-parabolic geometry may not be the case downstream, and a jet geometry break may occur near this region. However, when considering a smoothly connected broken power law to check this possibility, uncertainties were dramatically large and results were not reliable, possibly due to the significant oscillation in the jet radius from the University College Cork data. We note that this data correspond to a single epoch observation. Nonetheless, \cite{Pushkarev17} fitted the whole region as a geometrically conical jet, with $\epsilon=0.95\pm0.01$. This supports the possibility of a jet geometry break near the Bondi radius. This phenomenology has the features of the one seen in M87 \citep{Asada12}, NGC6251 \citep{Tseng16} or 3C~273 \citep{Akiyama18}. If this is the case, this would strongly support a unification in the jet geometry aspects for various kinds of AGN.

%Interestingly, it seems that the extrapolation of the jet width (dashed line in Figure \ref{fig:width_rg}) would not smoothly connect with the super massive black hole (SMBH), suggesting that the jet footpoint may be not anchored to the ergosphere. If this is the case, the footpoint may be anchored to the disk, which may suggest a centrifugally driven outflow \citep{BP82} mechanism for the jet origin. This scenario would also be in agreement with the indications of a particle dominated region for the new jet components ejection suggested in Paper~II. 

The extrapolation of the jet width (dashed line in Figure \ref{fig:width_rg}) suggests that the jet foot point is located at a certain distance of the SMBH (for a jet radius $=1~r_s$, the distance is $2.0\pm0.7~r_s$). If so, this may point towards a rotationally driven outflow \citep{BZ77} mechanism for the jet origin. This scenario would give rise to a leptonic jet with little matter load, and initially Poynting flux dominated, in tension with the results from Paper~II, which suggested a particle-dominated region for the new jet components ejection.

It is possible however that the trend analyzed here does not hold upstream the jet. On the other hand, as we mentioned above, our observations are reaching these regions where the jet opening angle is larger than the viewing angle and we are inspecting \emph{through} the jet, which would affect our discussion. This effect may be even more significant if the jet consists of a spine and sheath, as mentioned above.%  Alternatively, if the central engine consists of a double black hole system, as suggested by \cite{Liu10}, the interpretation towards the innermost regions may be different.

In order to check for this and the robustness of our results, the jet geometry needs to be scrutinized with data spanning more decades along the jet. Given that we need to probe towards the vicinity of the central engine, within the Bondi radius region, high resolution VLBI observations, with e.g., RadioAstron, the Event Horizon Telescope (EHT) \citep[e.g.,][]{Doeleman12,Lu13,Gomez16,EHTC2019} or future facilities and instruments such as Millimetron \citep{Kardashev14}, will be necessary.

%\subsection{Jet Evolution}
Based on a number of theoretical studies \citep[see e.g.,][]{Komissarov07,Komissarov09,Tchekhovskoy08,Lyubarsky09}, we expect to observe deviations from equipartition at the jet base that should be dominated by the magnetic field. This energy should be converted to kinetic energy as the bulk acceleration proceeds; however, this is not observed in 1633+382, at least not on the scales examined on this study. Unlike M87, the source shows collimation but not acceleration. Other studies \citep[e.g.,][]{Kino02,Homan06,Nokhrina17,Pilipenko18} have also proposed models where the energy density of particles dominates over that of the magnetic field in the innermost regions, as appears to be the case for 1633+382.  In order to investigate and understand this in more detail, sub-mm observations of this source will be useful.

%In Paper~II we discussed the flux density enhancement observed in 1633+382 due to the emergence of two new components, namely C2 and C3, from the VLBI radio--core. The evolution of turnover flux and turnover frequency follow the aspects of a shock--in--jet model. It was suggested there that the source of the flux injection, leading to the ejection of new components, showed a significant departure from equipartition. Other studies \citep[e.g.,][]{Kino02,Homan06,Nokhrina17,Pilipenko18} have also proposed several departure from equipartition conditions near the base of the jet in some sources. This suggests that the collimation and acceleration mechanisms are still at work and the magnetization parameter should generally decrease its value along the jet axis in exchange for jet acceleration.

%On the other hand, here we suggest that, at distances of the order of few mas ($10^{6-7} r_g$), close to the Bondi radius, the jet components show constant speeds. If this is the case, jet acceleration and collimation occur within the first $10^{5-6}r_g$, well within the Bondi radius. A scenario where acceleration occurs upstream the Bondi radius has already been suggested for e.g., M87 \citep{Asada14,Hada17}, although there appears to be a difference between M87 and 1633+382: while acceleration seems to hold down to the Bondi radius itself in M87, acceleration is not found at such scales in 1633+382. In order to investigate and understand in more details how this transition occurs, sub-mm observations of 1633+382 will be useful.

\subsection{Magnetic Field and Electron Densities}

Under an assumption of a jet with constant Lorentz factor of the emitting electrons and power-law dependences of the particle density, magnetic field strength and jet transverse size in the form $N_e\propto r^{-n}$,$B\propto r^{-b}$, and $R\propto r^{\epsilon}$, the brightness temperature for optically thin synchrotron emission should follow a power law \citep{Kadler04,Kravchenko16,Beuchert18}
\begin{equation}
T_{b,jet}\propto r^{-f}\quad;\quad f=-\epsilon+n+b(1-\alpha).
\label{eq_Tb}
\end{equation}

The parameter $n$ can be constrained a priori by the geometry of the outflow, since $N_e\propto R^{-2}$. This does not require any assumption about the physical conditions in the jet and only requires mass conservation. If we take $\epsilon=0.71\pm0.06$ and $f=2.3\pm0.1$, we find $n=1.4\pm0.1$. This leads to the relation $1.6=b(1-\alpha)$. For small values of the spectral index, this suggests $1<b<1.6$.

If we consider core shift effects, we can go a step further. In general, the position of the core is proportional to the frequency in the form $r_c(\nu)\propto \nu^{-1/k_r}$, where the coefficient $k_r$ holds information about physical conditions in the ultracompact jet region: 
\begin{equation}
k_r=\frac{(3-2\alpha)b+2n-2}{5-2\alpha}.
\label{eq_kr}
\end{equation}
In the same manner, under the assumption that each jet component is an independent relativistic shock with adiabatic losses dominating the emission, we can consider \citep{Marscher90}:
\begin{equation}
T_{b,jet}\propto d^{\xi}\quad;\quad      \xi=-[2(2s+1)+3b(s+1)]/6,
\label{eq_xi}
\end{equation}
where $s=1-2\alpha$ and $d$ is the component size. 
Thus, if we combine Equations \ref{eq_Tb}, \ref{eq_kr} and \ref{eq_xi}, and we can estimate the value of $k_r$, we can derive $b$ and $n$, together with the spectral index $\alpha$ without further assumptions. Several studies \citep[e.g.,][]{OSullivan09,Sokolovsky11,Zdziarski15} have shown that, in general, $k_r\sim1$, as expected from equipartition arguments. Although this is not always the case for all sources \citep[see e.g.,][]{Kadler04,Bach08,Kutkin14}, \cite{Algaba12} found $k_r=0.9\pm0.1$ for 1633+382, which is consistent to the expected equipartition value. These considerations lead to the values of power--law indices $b=1.5\pm0.2$ and $n=1.1\pm0.2$, with a spectral index of $\alpha=-0.25\pm0.03$. These values are in agreement with the tentative ones discussed above.

If we consider simple flux density conservation without significant energy dissipation, poloidal magnetic fields scale as $B \propto R^{-2} \propto r^{-1.4}$, while toroidal fields scale as $B \propto R^{-1} \propto r^{-0.7}$, where we have used $R \propto r^{0.7}$. The value of $b=1.5\pm0.2$ derived here thus suggest that poloidal magnetic fields are dominant in the jet. This implies that, under an assumption of a helical magnetic field distribution, the overall magnetic field in this region may maintain relatively larger pitch angle (e.g., $>45\degr$), or the emission region with larger pitch angle magnetic field (e.g., the spine structure, if a spine-sheath structure is considered) may dominate the radiation at radio wavelengths.

\subsection{The Origin of the Jet Structure}

Interestingly, the jet shows a significant bend within the first mas. Similarly, the components flux density seems to deviate from a pure power--law and to reach a local plateau within these distances (see Figure \ref{Flux}). It is possible that these two effects are related if the flux density plateau is produced by a change in the Doppler factor due to the jet bending. This bending may be caused by several different factors. One possibility could be interaction with the ambient medium. Indeed, in powerful FSRQs such as 1633+382, the presence of cold gas in the central regions is very likely, and not necessarily uniformly distributed. Within the Bondi radius, the central black hole should be having more influence and playing a much larger role, and ambient medium effects would potentially lead to the presence of pressure gradients which would translate into variations in the core shift offset $\Omega_{r\nu}$ \cite{Lobanov98}, which are not seen by investigating the values obtained from different frequency pairs based on the core shift for 1633+382 in \cite{Algaba12}. However, a significant amount of gas may be located downstream the jet. The jet bending could also result from the projection in the sky of a helical or harmonic jet trajectory, due to the presence of helical magnetic fields and/or of a binary SMBH.%Another possibilities include the projection of the jet helical or harmonic trajectory in the sky induced by helical magnetic fields and/or the influence of a binary SMBH as its central engine.

In order to investigate these possibilities further, we stacked different epochs from the BU Blazar monitoring program from 2007 to 2016 weighted by their rms noise, so that we can increase the dynamic range for our analysis. We take into consideration two things here: i) the position angle PA of the jet seems to be slightly changing over time \citep[see e.g., ][]{Ro19}, but the standard deviation of the PA for the epochs considered here is about 10\degr, which is acceptable for our purposes.  ii) Periods of high flux densities which may affect the overall morphology of the source. In order to check this, we performed a similar analysis stacking only the images corresponding to a quiescent period. We checked that the results do not significantly differ.

\begin{figure}
\center
\includegraphics[scale=0.60,trim={2.8cm 0cm 0cm 1cm},clip]{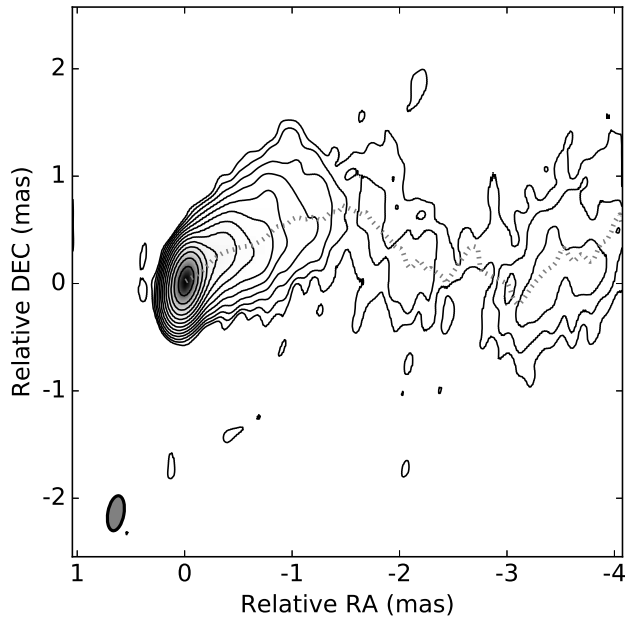}
\caption{Stacked map of BU Blazar monitoring program for 1633+382. The root mean square of the map is rms=0.14~mJy. Contours start at $3\times$rms and increase in steps of 2. The grey oval in the bottom left of the map indicates the beam size. Dotted lines follow the ridge line of the jet.}
\label{fig_stacked}
\end{figure}

The stacked map is shown in Figure \ref{fig_stacked}. With the improved rms levels, details of the jet are revealed beyond 1~mas and its structure can be well followed up to 4~mas from the core. The bending of the jet, which was suggested by the component model-fitting is now clearly seen, and another possible bending is suggested at around 3~mas. This figure compares well with the corresponding MOJAVE\footnote{\url{http://www.physics.purdue.edu/astro/MOJAVE}} images at 15~GHz, which clearly show the innermost jet bending, albeit the comparatively lower resolution, and the knot at 3-4~mas from the core. Overall, we find that the projection of the jet in the sky resembles that of a sinusoidal pattern.

We obtained the ridge line of the jet following \cite{Cohen15}. First, we made slices across the jet and fitted a gaussian to each profile. We considered both the peak of the fitted gaussian and the midpoint of the intensity across the jet as estimates for the ridge line. The points were thereupon smoothed with a third-order spline. Both methods showed consistent results and followed well each other, deviating at most of the order of $\sim$0.1~mas, which we will hereafter consider our uncertainty. The resulting ridge line is shown in Figure \ref{fig_stacked}. The apparent sinusoidal pattern that the overall jet structure seems to follow is confirmed by the ridge line analysis.

\begin{figure}
\center
\includegraphics[scale=0.45,trim={0cm 0cm 0cm 1cm},clip]{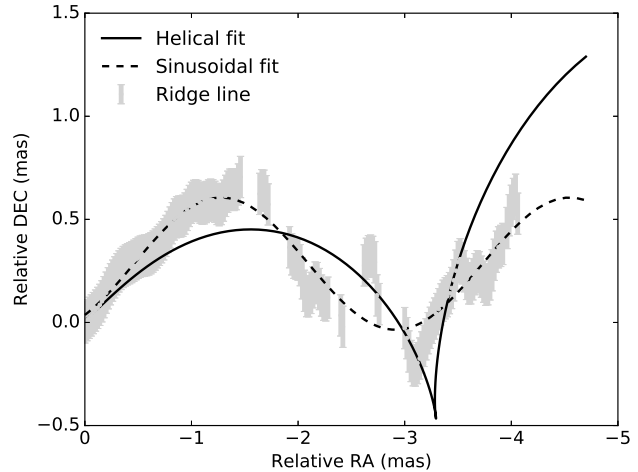}
\caption{Models considered for fitting the jet ridge line of the stacked maps of 1633+382. Straight line: projection of a helical semi-parabolic pattern; dashed line: sinusoidal pattern.}
\label{fig_helical_fitting_to_ridgeline}
\end{figure}

We consider now the possible origin for this morphology. As mentioned above, one of the possibilities arises as the projection of a helical pattern. We follow \cite{Kun14} to model the jet with a helical structure. Note that the conical helix they considered for 1928+738 is not appropriate for 1633+382, as we just found out that the jet geometry appears to be quasi-parabolic in the regions of our concern. Hence, the modified parametrization of the projected helical structure in the plane of the sky will become
%\begin{align}
%x^{\rm proj}_{\rm jet}&=a_1 u^{\epsilon} \cos{u}\cos{\theta}+a_2 u \sin{\theta} \nonumber \\
%y^{\rm proj}_{\rm jet}&=a_1 u^{\epsilon} \sin{u}
%\end{align}

\begin{align}
x^{\rm proj}_{\rm jet}&=F(u)\cos(\lambda) - G(u) \sin(\lambda) \nonumber \\
y^{\rm proj}_{\rm jet}&=F(u)\sin(\lambda) + G(u) \cos(\lambda)
\end{align}

where 

\begin{align}
F(u)&=b u^{\epsilon} \cos({u-\phi})\cos{\theta}+a u \sin{\theta} \nonumber \\
G(u)&=b u^{\epsilon} \sin({u-\phi})
\end{align}

with $u$ the polar angle measured in the plane perpendicular to the jet axis, $\phi$ an initial phase, $\lambda$ the P.A. rotation angle, $\epsilon=0.71$ and $\theta=2^{\circ}$. In Figure \ref{fig_helical_fitting_to_ridgeline} we show the best model, which is found for $a=21.0\pm0.5$, $b=0.25\pm0.05$, $\phi=6.0\pm0.2$~rad and $\lambda=0.08\pm0.04$~rad. The model does not to match well with several features of the ridge line: the first bend of the jet appears significantly sharper than expected by the model, and the trajectory downstream the jet seems not to follow the direction of the model. Furthermore, if the jet were to follow a helical pattern, this could produce a signature in the observed flux density due to Doppler boosting drifting as the viewing angle of the jet changes in its helical path. Although there is a flux density plateau which deviates from a single power law shown in Figure \ref{Flux}, this feature does not correlate with the ridge line pattern. Instead, the apparent kink at about 3.5~mas, for example, may be related with the properties of the observed jet knot at approximately that location.

%Furthermore, the flux density plateau and, in particular, the residuals of a power law fitted to the flux density shown if Figure \ref{Flux}, do not seem to follow the fitted helical curve. This indicates that such flux plateau seems not to be simply originated from Doppler boosting drifting due to variation in the viewing angle, as previously speculated. We note however the apparent kink at about 3.5~mas, which may be related with the properties of the jet observed knot at approximately such location.

Another appealing possibility is that the jet is precessing with time and the change of the position angle of the jet with distance is an indicator of such precession. Indeed, although we mentioned earlier that the innermost jet position angle does not change much during the period discussed here, \cite{Ro19} suggests that, over much larger periods of time, there is a trend of the innermost jet position angle to change. If this is the case, we can estimate a possible precession period based on the sinusoidal pattern of the jet together with the jet speed. We consider a fit of the form 
\begin{equation}
y^{\rm proj}_{\rm jet}=A\sin(Bx^{\rm proj}_{\rm jet}+C)+D,
\end{equation}
which adjusts to the data with $B=1.92\pm0.02$~mas. Combined with a speed of 140~$\mu$as yr$^{-1}$ (see Paper~II), this suggests a precession period of $T_{obs}\sim23\pm5$~yr.  \cite{Ro19} also suggest that, if there is a periodic swinging of the jet, the period may be larger than $\sim20$~years. The precession model, shown with a dashed line in Figure \ref{fig_helical_fitting_to_ridgeline}, also provides a better fitting for the data, with the ratio of  $\chi^2$ of the order of $\chi^2_{\rm helical}/\chi^2_{\rm sinusoidal}\sim7$, although this alone is not sufficient to prefer any model. It is quite possible that the curving and wiggling of the jet results as a combination of different factors, including precession, helicity and instabilities or the growth of magneto-hydrodynamic instabilities. Some of such would include Kelvin-Helmholtz  \citep[see e.g.][]{Perucho06} or current-driven \citep[see e.g.][]{MeierNakamura06} instabilities. Other models where external causes, such as interaction with the surrounding medium, used to explain the wiggling, should also be considered.

\section{Conclusions}
In this paper we focus on the properties of the radio jet of 4C~38.41 (1633+382) and the evolution of the emerged components associated with the observed flux enhancement. For this purpose, we use high--resolution 43~GHz images from the  BU-VLBA  Monitoring program and study multi--epoch properties of the jet in terms of jet components. We consider the trajectory of the innermost regions of the jet, as well as its size and brightness temperature, and its relation with distance from the VLBI radio--core.

We find that the jet width follows a semi-parabolic geometry, with $R\propto r^{0.7}$, between the pure conical and the parabolic cases within regions considered to be upstream the jet, within mas scales. Additionally, there are hints suggesting the geometry downstream the jet may be conical, which indicates a possible collimation break at distances of the order of $10^6~r_g$, possibly within scales not too different from the estimated Bondi radius. This result is similar to that found in other sources, strengthening the suggestion that this may be a global characteristic of AGN jets. 

Considering the brightness temperature profile and the geometry, together with core shift arguments, we can estimate the power--law dependences on the electron density and magnetic field strength to be in the form  $N_e\propto r^{-1.1\pm0.2}$ and $B\propto r^{-1.5\pm0.2}$, respectively, for the outmost components and the general trend. This suggests that the magnetic field is predominantly poloidal in the regions discussed here.

We analyse the structure of the innermost jet revealed by image staking. The innermost jet trajectory does not appear to be ballistic and, instead, matches well with a sinusoidal pattern. We consider various models from where the pattern arises, including a helical pattern or jet precession. %Our data is compatible with i) a model where the central engine consists of a binary black hole with mass ratio $\nu\sim0.3$, separated by a projected distance of about $3~\mu$as, and the change in the direction of the jet is produced due to the spin precession of the dominant SMBH, or ii) a Lense-Thirring effect due the of the precession of the accretion disc over a fast rotating SMBH.

\acknowledgments
\footnotesize{\emph{Acknowledgements.}
%We are grateful to all staff members in KVN who helped to operate the array and to correlate the data. The KVN is a facility operated by the Korea Astronomy and Space Science Institute. The KVN operations are supported by KREONET (Korea Research Environment Open NETwork) which is managed and operated by KISTI (Korea Institute of Science and Technology Information). 
The authors are very grateful for a very fruitful discussion with M. Nakamura and H. Ro. This study makes use of 43 GHz VLBA data from the VLBA-BU Blazar Monitoring Program (VLBA-BU-BLAZAR), funded by NASA through the {\it Fermi} Guest Investigator Program. The VLBA is an instrument of the National Radio Astronomy Observatory. The National Radio Astronomy Observatory is a facility of the National Science Foundation operated by Associated Universities, Inc. 
%This research has made use of data from the OVRO 40-m monitoring program (Richards, J. L. et al. 2011, ApJS, 194, 29) which is supported in part by NASA grants NNX08AW31G, NNX11A043G, and NNX14AQ89G and NSF grants AST-0808050 and AST-1109911. 
%This work used Submillimeter Array data. The Submillimeter Array is a joint project between the Smithsonian Astrophysical Observatory and the Academia Sinica Institute of Astronomy and Astrophysics and is funded by the Smithsonian Institution and the Academia Sinica. 
%G. Zhao is supported by Korea Research Fellowship Program through the National Research Foundation of Korea (NRF) funded by the Ministry of Science, ICT and Future Planning (NRF-2015H1D3A1066561). 
%J. C. Algaba, D.-W. Kim and S. Trippe 
J. C. Algaba acknowledges support from the National Research Foundation of Korea (NRF) via grant NRF-2015R1D1A1A01056807 and support from Fundamental Research Grant Scheme (FRGS) FRGS/1/2019/STG02/UM/02/6. J. Park acknowledges support from the NRF via grant 2014H1A2A1018695. S. S. Lee was supported by the National Research Foundation of Korea (NRF) grant funded by the Korea government (MSIP) (No. NRF-2016R1C1B2006697). M. Kino acknowledges JSPS KAKENHI Grant Numbers JP18K03656 and JP18H03721. This research was supported by an appointment to the NASA Postdoctoral Program at the Goddard Space Flight Center, administered by Universities Space Research Association through a contract with NASA. }

%This research has made use of data from the MOJAVE database that is maintained by the MOJAVE team \citep{Lister09}. This research has made use of the NASA/IPAC Extragalactic Database (NED) which is operated by the Jet Propulsion Laboratory, California Institute of Technology, under contract with the National Aeronautics and Space Administration. The National Radio Astronomy Observatory is operated by Associated Universities, Inc., under contract with the National Science Foundation.

%{\it Facilities:} \facility{VLA}.

\end{document}